\documentclass{article}
\usepackage{amsmath}
\usepackage{amsfonts}
\usepackage{amssymb}
\usepackage{tikz}
\usepackage{graphicx}
\usepackage{lastpage}
\usepackage[margin=1.0in]{geometry}
\usepackage{fancyhdr, lastpage}
\usepackage{float}
\usepackage{epsfig}
\usepackage{tabularx}
\usepackage[font=normalsize,labelfont=bf]{caption}
\usepackage{xcolor}
\usepackage{titling}
\usepackage{cite}
\usepackage{soul}
\usepackage{cancel}
\usepackage{subcaption}
\usepackage{bm}

\usepackage[colorlinks = true,
            linkcolor = blue,
            urlcolor  = blue,
            citecolor = blue,
            anchorcolor = blue]{hyperref}

\begin{document}
\title{\vspace{-1.15cm}\textbf{Zonal flow suppression of turbulent transport in the optimized stellarators W7-X and QSTK} \vspace{-2.5em}}
\date{}
\maketitle
\vspace{-1.5cm}
\begin{center}
\fontsize{10}{15}\selectfont Abhishek Tiwari$^{1,*}$, Joydeep Das$^{1}$, Jaya Kumar Alageshan$^{1}$, Gareth Roberg-Clark$^{2}$, Gabriel Plunk$^2$, Pavlos Xanthopoulos$^2$, Sarveshwar Sharma$^{3,4}$, Zhihong Lin$^{5}$ , Animesh Kuley$^{1,*}$  \\
\vspace{0.25cm}
\small \textit{$^{1}$Department of Physics, Indian Institute of Science, Bangalore 560012, India} \\
\small \textit{$^2$Max-Planck-Institut für Plasmaphysik, D-17491 Greifswald, Germany} \\
\small \textit{$^3$Institute for Plasma Research, Bhat, Gandhinagar 382428, India} \\
\small \textit{$^4$Homi Bhabha National Institute, Anushaktinagar, Mumbai, Maharashtra 400094, India}\\
\small \textit{$^5$Department of Physics and Astronomy, University of California Irvine, CA 92697, USA}\\
\textit{$^{*}$Email: abhishektiwa@iisc.ac.in, akuley@iisc.ac.in}\\
\end{center}

\begin{abstract}
\noindent We present a comparative study of transport in two optimized stellarator configurations: Wendelstein 7-X (W7-X) and a recent design called Quasi-Symmetric Turbulence Konzept (QSTK). Using global Gyrokinetic Toroidal Code (GTC), we explore the role of zonal flows (ZFs) in suppressing electrostatic Ion Temperature Gradient (ITG) driven turbulence in both configurations. The simulations reveal that ZFs significantly reduce ion heat transport in both W7-X and QSTK, with a   lower value of heat flux on the latter configuration, as suggested by the apparently higher linear threshold (``critical'') gradients for ITG modes. The study also highlights that both stellarators exhibit similar mode structures. The results support the notion that linear stability measures, in combination with nonlinear stabilization by zonal flows, can play an important role in the suppression of nonlinear heat fluxes.
\end{abstract}


\section{\fontsize{12}{15}\label{sec:intro}Introduction}

\noindent Recent advancements in stellarator\cite{ Velasco,LION2025114868} design and technology have significantly improved their plasma confinement capabilities, rendering them an increasingly promising approach in fusion research alongside the well-established tokamak designs. The stellarator has advantages over the tokamak, for instance, the toroidal 
current, steady state operation, and lower magnetohydrodynamic (MHD) activity. However, these advantages come at the cost of breaking toroidal symmetry, which can lead to an increase in collisional transport, 
coupling of macro- and micro-instabilities, and stronger damping of zonal 
flows\cite{Nicolau_2021,mishchenko2008collisionless}. Design and 
optimization\cite{Beidler2021} of stellarators have led to better plasma confinement in 
cases like W7-X\cite{Dinklage2018}, which have achieved a performance closer to 
tokamaks. It has been 
confirmed that neoclassical transport in W7-X is reduced with respect to non-optimized 
stellarators\cite{Carralero_2021,Beidler2021}.  However, turbulence has  
played a dominant role in limiting plasma performance in W7-X for specific heating scenarios.\cite{Beurskens_2021}.

\noindent A primary obstacle in plasma confinement is the excitation of micro-instabilities such as the ion temperature 
gradient (ITG) and the trapped electron mode (TEM). Turbulence associated with these drift wave instabilities can degrade
plasma confinement by transporting energy and particles. In modern stellarator experiments such 
as W7-X, advanced diagnostic techniques like phase contrast imaging 
(PCI) are employed to measure and characterize ITG and TEM behaviour\cite{10.1063/1.5038804}. It 
has been found that the stability of the ITG 
mode depends upon the gradient ratio $\eta_i =L_{n_i}/L_{T_i}$, where $1/L_X=-(1/X)(dX/dr)$ is the 
gradient length scale. The critical gradient (CG) is the threshold gradient 
for the onset of the ITG mode. One way to combat losses from ITG is to increase the size and heating 
power of the stellarator\cite{PhysRevResearch.5.L032030}. Another way is to address ITG itself by changing the plasma profiles\cite{podavini2024iontemperaturedensitygradient}. Also, radio frequency waves can be used to stabilize these micro-instabilities in fusion plasmas\cite{Kuley09,Kuley10}. 

\noindent In addition to these strategies, shaping the magnetic field can further reduce losses due to micro-turbulence in the plasma core. Certain implementations of this strategy target the critical gradient of the 
mode\cite{PhysRevResearch.5.L032030}, (CG-approach), producing the HSK stellarator, which exhibits the most significant critical gradient at half radius of all known stellarators. It has also been shown in Ref.~\cite{PhysRevResearch.5.L032030} that this strategy can target the CG of the toroidal branch of the ITG mode without compromising MHD stability. Such optimization produces a quasi-helical symmetric 
configuration (QSTK) with strongly reduced ITG turbulence and acceptable levels of neoclassical 
losses, and MHD stability, leading to improved ion confinement.

The Quasi-Symmetric Turbulence Konzept (QSTK) configuration has $N_{fp}=6$,
an aspect ratio of 7.5, neoclassical transport coefficient $\epsilon_{\text{eff}} < 1 \%$ up to 
half radius, large rotational transform $>1.6$ and $\simeq 5\%$ alpha particle neoclassical
losses for particles initialized at half radius. QSTK also features good MHD stability, small bootstrap current, and it admits
coils of moderate complexity. In addition,  flux-tube based gyrokinetic simulations suggest that
 the heat flux is significantly reduced compared to W7-X \cite{PhysRevResearch.5.L032030}.

\noindent Several gyrokinetic simulations of micro-turbulence in stellarators have been done 
previously. For example, the global code EUTERPE\cite{KLEIBER2024109013,Riemann_2025} was used to study the effects 
of radial electric field on linear ITG instability in W7-X and LHD\cite{Riemann_2016}. The 
effect of density gradient and micro-instabilities on turbulent heat transport in stellarators 
was performed with flux-tube code \texttt{stella}\cite{thienpondt2024influencedensitygradientturbulent}. The 
electromagnetic gyrokinetic Vlasov flux-tube code GKV was used to study the impact of isotope ion mass on
TEM driven turbulence and zonal flow in LHD stellarators\cite{PhysRevLett.118.165002}. The 
GENE flux-tube simulation has been used to study the effect of ZF dynamics and turbulent transport in stellarator
geometry\cite{10.1063/1.3560591}. 
The codes GENE-3D, KNOSOS, and TANGO were used to compute the plasma 
profiles due to the combined effect of neoclassical transport, turbulent transport, and external 
particles in W7-X, QSTK and HSK stellarators\cite{Navarro_2023,navarro2023assessingglobalionthermal}. The GT5D code performed full-f global simulations in LHD and the collisionless zonal flow damping\cite{10.1063/1.5010071}. Global XGC-S\cite{10.1063/1.5140232} and GENE-3D\cite{Navarro_2020} were used to carry out 
micro-turbulence simulations using adiabatic electrons in W7-X and LHD. The global gyrokinetic toroidal code GTC was used to perform the nonlinear global gyrokinetic simulations of micro-turbulence in LHD and W7-X, including the kinetic effect of electrons in stellarators\cite{Singh_2022,Singh_2024,Nicolau_2021}. In addition to these efforts, global fluid simulation of plasma turbulence in 
stellarators has been carried out using the GBS \cite{Coelho_2024} and BSTING \cite{Shanahan_2019, Shanahan_Bold_Dudson_2024} codes. 

\noindent In previous works \cite{Singh_2022,Singh_2024,Nicolau_2021}, the global gyrokinetic toroidal code GTC has been used to perform nonlinear global gyrokinetic simulations of micro-turbulence in LHD and W7-X, including the benchmark of ITG simulations with EUTERPE\cite{WangPoP20}, the suppression of ITG by neoclassical ambipolar electric field and its effects on microturbulence in W7-X stellarator\cite{FuPoP21}, the isotope effects\cite{Qin_2024}, the geometry effect on zonal flow\cite{chen2025geometry}, and the kinetic effect of electrons in stellarators. In this paper, we compare the effect of zonal flow on the turbulent transport driven by ITG turbulence with adiabatic electrons for the stellarators W7-X and QSTK. Recently, Carralero et al, have shown agreement between the experimental observations and the gyrokinetic simulation of low-frequency ZF using local (Stella) and global (EUTERPE) codes \cite{Carralero}. Also, it's important to mention that these simulations are very long and focused explicitly on resolving very low-frequency (sub-kHz) oscillation to compare specifically with the experimental measurements. The paper is organized as follows: First, in Sec.~\ref{sec:simmodel}, we briefly present the physics model and the numerical code employed. Then, in 
Sec.~\ref{sec: w7xqstkinst}, we study the linear simulation of ITG in both stellarators. 
In Sec.~\ref{sec:nlsimstels}, we perform the nonlinear simulations and evaluate the effect of zonal flow on the ITG turbulence.
We conclude with some discussion in Sec.~\ref{sec: conclude}.

\section{\fontsize{12}{15}\selectfont \label{sec:simmodel} Simulation Model}

\noindent In this paper, we use the global nonlinear code  GTC\cite{Lin1998} to perform collisionless gyrokinetic simulations of micro-turbulence. GTC has been extensively applied to simulate neoclassical and turbulent transport\cite{Xiao2009,Singh_2023,Singh_2024,Singh_2024b,singh2025}, Alfv\'en waves\cite{PhysRevLett.111.145003,liu2022regulation}, energetic particles\cite{Wenlu2008,Brochard2024}, and radio frequency waves\cite{Kuley2013,Kuley2015,Bao_2016} in toroidally confined plasmas. 

\begin{figure}
\includegraphics[width=0.7\textwidth]{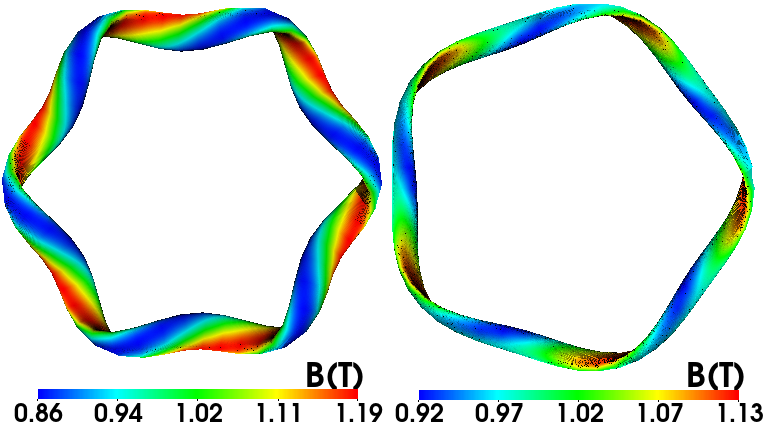}
\centering
\caption{\label{fig:qstkconfig}The magnetic field of the two
stellarators: (Left) QSTK; (Right) W7-X on the flux surface with $\psi/\psi_w=0.57$ ; the colors represent the corresponding strength of the magnetic fields.}
\end{figure}

\noindent GTC interfaces with VMEC\cite{osti_5537804}, an ideal MHD code, to obtain the non-axisymmetric equilibrium of QSTK and standard W7-X configuration, considering closed magnetic surfaces. This equilibrium data contains information on poloidal current, toroidal current, and magnetic field described as Fourier series in poloidal and toroidal direction, given by,
\begin{eqnarray}
\mathcal{F}(\psi,\theta,\zeta)= \sum_n \; \left[ \mathcal{F}_c(\psi,\theta,n) \; \cos(n\zeta) + 
     \mathcal{F}_s(\psi,\theta,n) \; \sin(n\zeta) \; \right] \nonumber
\label{eq:one}
\end{eqnarray}
\noindent where $(\psi,\theta,\zeta)$ are the poloidal flux, poloidal angle and toroidal angle, respectively. Here, n is the toroidal harmonic number and $\mathcal{F}_c$ and $\mathcal{F}_s$ are the Fourier 
coefficients specified on rectangular equilibrium mesh on the 
$\zeta=const.$ poloidal plane. 
GTC uses a global field-aligned mesh in real-space coordinates, which is used to represent all turbulence quantities. This provides computational efficiency without imposing any geometrical approximations and only needs a small number of grid points in the parallel directions to resolve the drift-wave eigenmode structure due to the anisotropic nature of  micro-turbulence. Due to the toroidal asymmetry in stellarators, more toroidal grid points are required than tokamak for generating 3D equilibrium quantities. In GTC, we use extra grid points between every two turbulence grid points for generating the equilibrium mesh, on which equilibrium magnetic fields are calculated and used to push particles. Also note that the QSTK has 6 field periods ($N_{fp}=6$), while the W7-X stellarator has 5 field periods ($N_{fp}=5$) i.e., all equilibrium quantities have a periodicity of $2\pi/N_{fp}$ in the toroidal direction. Therefore, we have constructed the spline on the equilibrium mesh for a field period of $\zeta=[0,2\pi/N_{fp}]$ with the toroidal periodicity explicitly enforced at $\zeta=0$ and $2\pi/N_{fp}$. We exploit the $2\pi/N_{fp}$ periodicity, to simulate one period of each configuration, instead of the full torus.

\noindent In the present work, we assume that the electrons follow a Boltzmann distribution. The 
collisionless gyrokinetic Vlasov equation, which describes the thermal ions in an inhomogeneous magnetic field, is given by\cite{Singh_2022,Singh_2024,brizard2007foundations}
\begin{eqnarray}
    \frac{d}{dt}f(\textbf{X},\mu,v_{||},t) = \left[ \frac{\partial}{\partial t}+ 
    \dot{\textbf{X}} \cdot \nabla + \dot{v}_{||}\frac{\partial}{\partial v_{||}} \right]f= 0;
    \quad\quad \dot{\textbf{X}} \; = \; v_{||}\textbf{\textit{b}} + \mathbf{v_d}+ \mathbf{v_E} 
    \label{eq:gyrovlas}
\end{eqnarray}
\begin{align}
\text{where,}\quad\quad\quad 
\dot{v}_{||} \; &= \; -\frac{1}{m} \frac{\bm{B^*}}{B} \cdot (\mu\nabla B + Z_i\nabla\phi) \nonumber \\
\mathbf{v_d} \; &= \; \frac{v_{||}^2}{\Omega}\; (\nabla \times \textbf{\textit{b}}) + \frac{\mu}{m\Omega} \; (\textbf{\textit{b}} \times \nabla B) \; \nonumber\\ 
\mathbf{v_E} \; &= \; \frac{c}{B} \; (\textbf{\textit{b}} \times \nabla \phi) \; \nonumber\\ \nonumber
\end{align}
where $f(\textbf{X},\mu,v_{||},t)$ is the particle distribution function, with $\textbf{X}$ is the gyrocenter position, $\mu$ is the  magnetic moment, $v_{||}$ is the parallel velocity, $Z_i$ is the ion charge, $m$ is ion mass, and $\phi$ is the electrostatic perturbed potential. $\bm{B}$ is the equilibrium magnetic field at the particle position, $\bm{B^*}= \bm{B} +\frac{Bv_{||}}{\Omega} \nabla \times \bm{b}$, and $\bm{b}= \frac{\bm{B}}{B}$. In the present work, we retain the zonal flow generated by the ITG turbulence, while neglecting the equilibrium radial electric field.

\noindent To reduce the particle noise in the simulation, GTC uses the $\delta f$ 
method\cite{10.1063/1.860870}. In this scheme, we decompose the distribution function into an
unperturbed equilibrium part and a perturbed part as $f=f_0+\delta f$. Further, the propagator
in  Eq.(\ref{eq:gyrovlas}) can be separated into an equilibrium part $L_0$ and a
perturbed part, $\delta L$ so that the Eq.(\ref{eq:gyrovlas}) can be written
as $(L_0+\delta L)(f_0 +\delta f)=0$, where
\[L_0= \frac{\partial}{\partial t}+ (v_{||}\textbf{b} + \mathbf{v_d})\cdot \nabla -\frac{1}
{m} \frac{\bm{B^*}}{B} \cdot (\mu\nabla B)\frac{\partial}{\partial v_{||}},\]
\[\delta L= \mathbf{v_E}\cdot\nabla -\frac{1}{m}\frac{\bm{B^*}}{B}\cdot Z_i\nabla\phi\frac{\partial}
{\partial v_{||}}\]
The equilibrium distribution function $f_0$ is determined by the condition $L_0 f_0=0$. The 
solution of this equation is approximated to be the local 
Maxwellian
\[f_{0}= \frac{n_{i}}{(2\pi T_{i}/m)^{3/2}} \exp\left( -\frac{2\mu B +mv_{||}^2}{2 T_{i}} \right)\]
where $n_i$ and $T_i$ are the equilibrium ion density and temperature, respectively. Next, we define the particle weight $w= \delta f/f_{0}$, and the evolution of this dynamical variable corresponding to thermal ions is given by
\begin{eqnarray}
    \frac{dw}{dt}= (1-w) \left[ -\mathbf{v_E}\cdot \frac{\nabla f_0}{f_0} + 
    \frac{Z_i}{m f_0} \frac{\bm{B^*}}{B} \cdot \nabla \phi \frac{\partial f_0}{\partial v_{||}}
    \right]
    \label{eq:dynamicalweightevolve}
\end{eqnarray}
\noindent We note from Eq.(\ref{eq:dynamicalweightevolve}) that we have neglected the neoclassical effect, since the term $\mathbf{v_d}\cdot\nabla f_0$ does not appear in the above equation. The electrostatic potential $\phi$ is obtained from the following gyrokinetic Poisson equation\cite{Singh_2023,Singh_2024b,lee1987gyrokinetic}, 
\begin{eqnarray}
    \phi-\tilde\phi = \frac{T_i}{n_i Z_i^2} (Z_i \bar{n}_i-en_e),
    \label{eq:gyropoisseqn}
\end{eqnarray}
where $\tilde\phi$ is the second gyro averaged potential, $\bar{n}_i$ and $n_e$ are the ion 
and electron guiding center density, respectively. In GTC, we can decompose the electrostatic potential $\phi$ and ion density perturbation $\bar{n}_i$ into zonal and non-zonal components as
\begin{eqnarray*}
    \phi= \langle\phi\rangle+ \delta\phi, \\
    \bar{n}_i= \langle \bar{n}_i \rangle+ \delta \bar{n}_i,
\end{eqnarray*}
with $\langle \delta \phi\rangle=0$, $\langle \delta \bar{n}_i\rangle=0$, $\langle\delta n_e\rangle=0$ and the $\langle \cdots \rangle$ represent flux-surface averaging. The non-zonal part of 
gyrokinetic Poisson equation thus becomes
\begin{eqnarray}
    &\delta\phi-\delta\tilde\phi= \frac{T_i}{n_i Z_i^2}(Z_i \; \delta \bar{n}_i-e \; \delta n_e)
    \; ; \\
    &\delta\tilde\phi=\frac{1}{2\pi}\int d^3\mathbf{v} \; \int d^3\mathbf{X} \; f_0(\mathbf{X}) \; \delta 
    \bar\phi(\mathbf{X}) \; \delta(\mathbf{X}+\mathbf{\rho}-\mathbf{x})
    \label{eq:nzgyropoiss}
\end{eqnarray}
where $\delta n_e=n_{0e}\delta\phi/T_e$, $T_e$ is the electron temperature. $\mathbf{x}$ and $\mathbf{X}$ are the particle position and the particle guiding center position coordinates, respectively, and $\mathbf{\rho}$ is the gyro-radius vector.
$\delta\bar\phi$ is the first gyro-averaged perturbed potential given by
\begin{eqnarray*}
    \delta \bar\phi(\mathbf{X})= \int d^3\mathbf{x}\int \frac{d\alpha}{2\pi} 
    \; \delta\phi(\mathbf{x}) \; \delta(\mathbf{x}-\mathbf{X}-\mathbf{\rho}),
\end{eqnarray*}
where $\alpha$ denotes the gyro-phase. Similarly, the ion perturbed density at the location of the guiding center is given by
\begin{eqnarray*}
    \delta \bar n_i(\mathbf{x})= \int d^3\mathbf{X}\int \frac{d\alpha}{2\pi} 
    \; \delta f(\mathbf{X})\;\delta(\mathbf{x}-\mathbf{X}-\mathbf{\rho}).
\end{eqnarray*}
\noindent A finite difference method is used to obtain the non-zonal electrostatic potential, while the flux-surface average gyrokinetic equation for the zonal component of electrostatic potential is computed using traditional integration in GTC.


\section{\fontsize{12}{15}\selectfont \label{sec: w7xqstkinst} Linear Simulation of ITG in stellarators}

\subsection{\label{subsec: w7xinst } ITG instability in W7-X}
\begin{figure}
\includegraphics[width=0.6\textwidth]{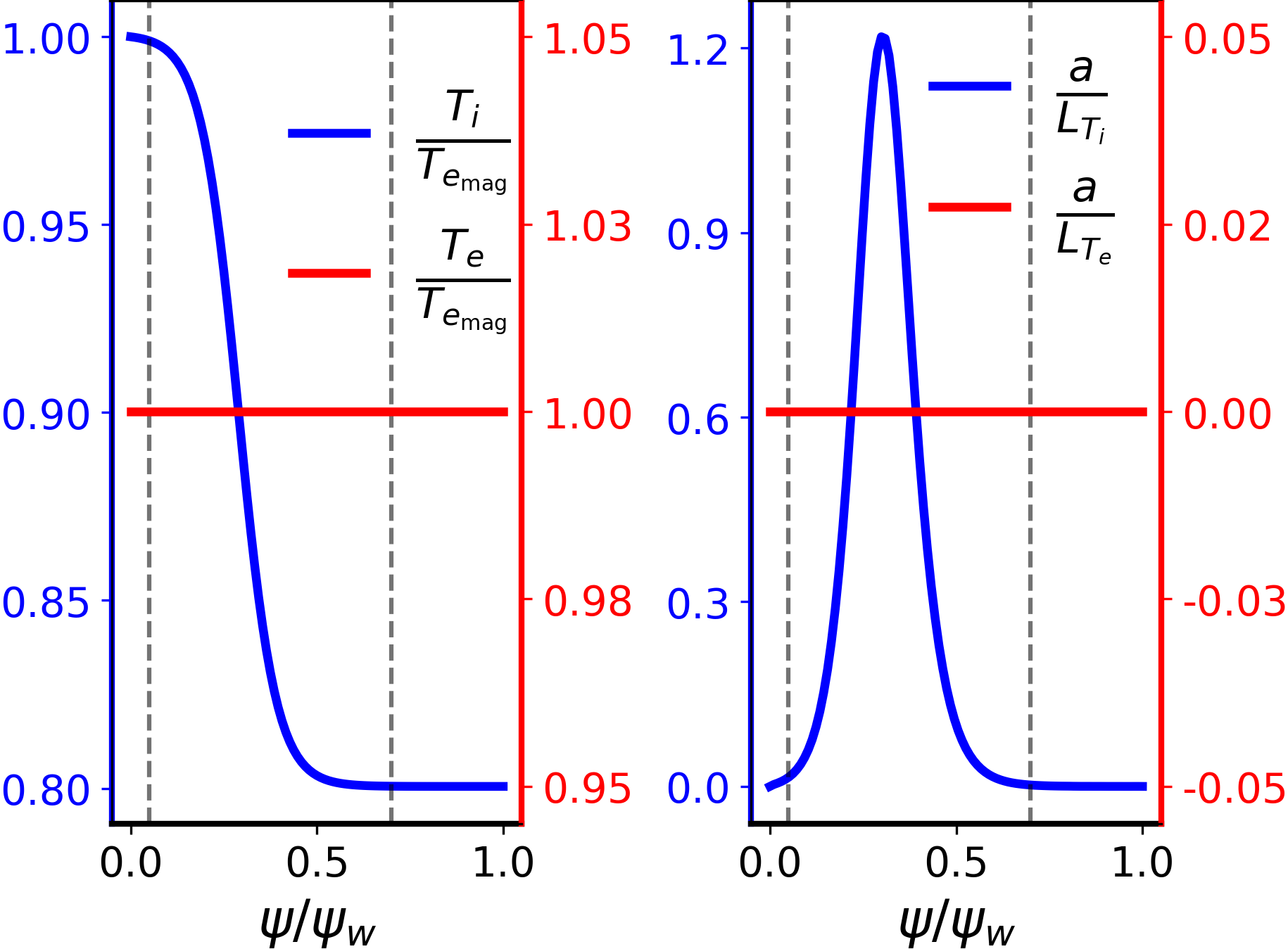}
\centering
\caption{\label{fig:tiprof}(Left) Radial profiles of equilibrium ion (blue) and electron (red) temperatures. Both quantities are normalized by $T_{e_{\text{mag}}}$, the electron temperature on-axis. (Right) We plot the quantity $a/L_{T_m}$ \{m=\{\textbf{i}ons,\textbf{e}lectrons\}\} as defined in Eq.~(\ref{eqn: tmprofile}). The dashed vertical lines indicate the simulation domain.}
\end{figure}

\begin{figure}
    \includegraphics[width=0.6\linewidth]{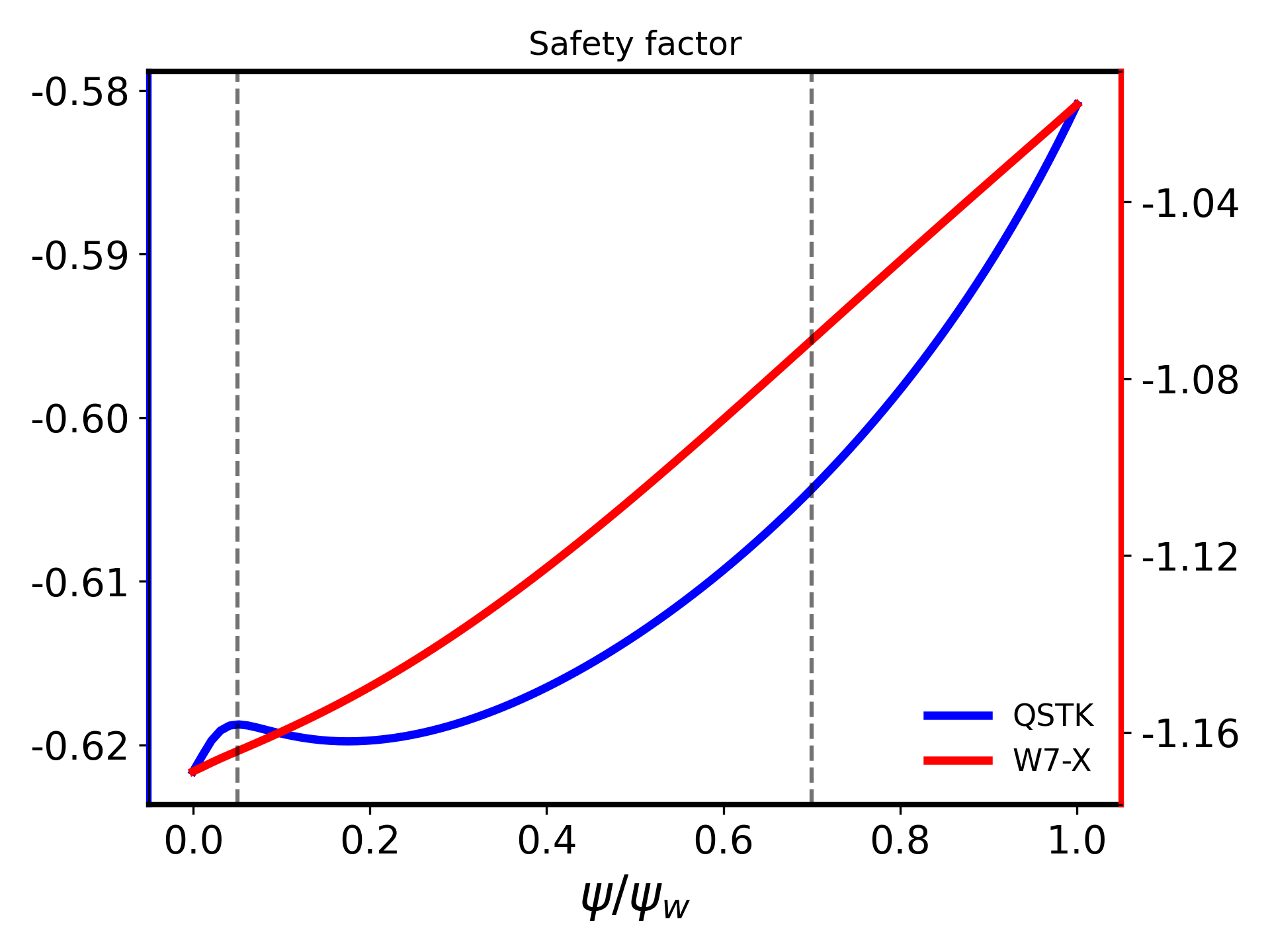}
    \centering
    \caption{Safety factor (q) for both W7-X (red) and QSTK (blue)  are
    shown in continuous curve and the dashed lines indicate the simulation domain.}
    \label{fig:safetyfactorstel}
\end{figure}

\noindent We apply the same plasma profiles for both QSTK and W7-X to simulate the linear and nonlinear physics of ITG turbulence in the two optimized stellarators and the effect of zonal flow. In Fig.~(\ref{fig:tiprof}) and Fig.~(\ref{fig:safetyfactorstel}) we show the temperature profile and safety factor, respectively, for W7-X and QSTK. The ion density $n_i$, electron density $n_e$, and electron temperature $T_e$ are assumed to be constant along radius, i.e., $\eta_i=\infty$. The definition of radial coordinate is $r=a\sqrt{\psi/\psi_w}$, with $a$ the minor radius corresponding to $\psi_w$. The temperature gradient length scale, measured relative to the minor radius $a$, is defined as 

\begin{eqnarray}
   \left(\frac{a}{L_{T_m}}\right)= -2\frac{\partial\ln T_m}{\partial \tilde{\psi}} \sqrt{\tilde{\psi}}  , \; \; \text{where} \; \; \frac{1}{L_{T_m}}= -
\frac{\partial \ln T_m}{\partial r} 
\label{eqn: tmprofile}
\end{eqnarray}
Here, $\tilde{\psi}= \psi/\psi_w$, with $m=\{i,e\}$ and $\psi_w$ represents the flux at the last close flux surface.
\begin{figure*}
  \centering \includegraphics[scale=0.6]{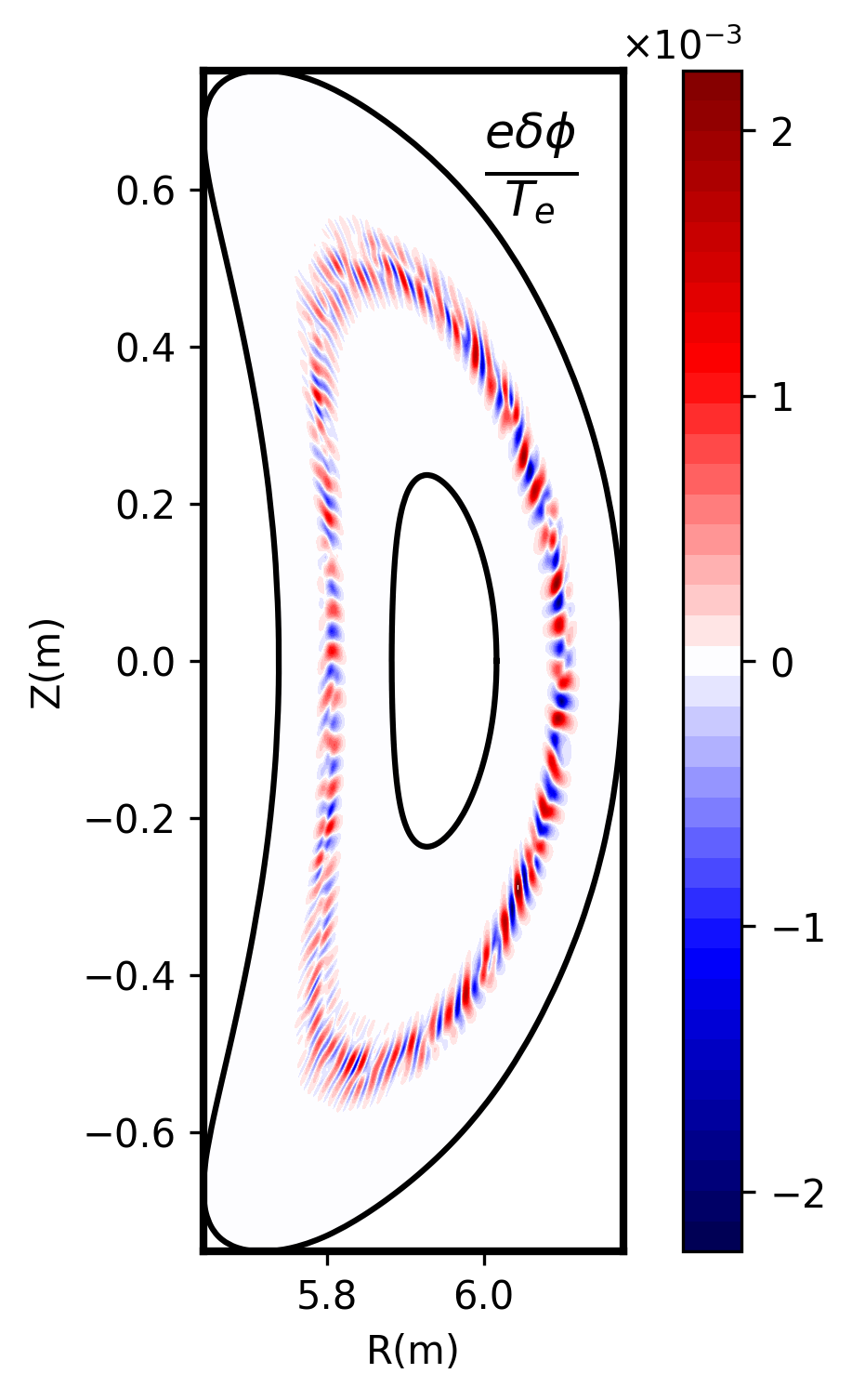}
   \centering \includegraphics[scale=0.6]{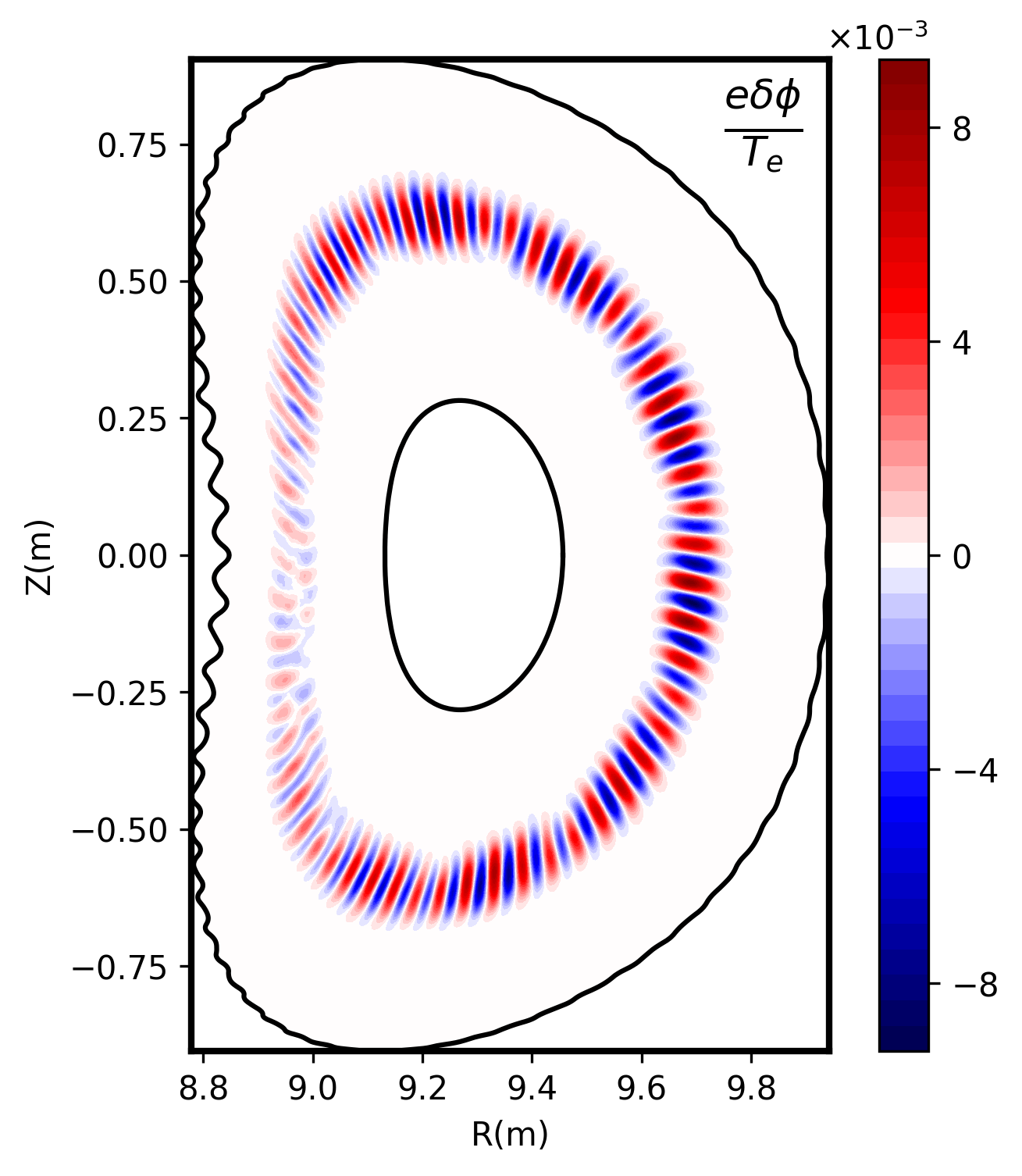}
  \caption{\label{fig:linmodew7x} The normalized electrostatic perturbed potential, $e\delta \phi/T_e$, on the $\zeta=0$ poloidal plane in the linear phase, at $t= 25.0R_0/C_s$ for W7-X (left) and at $t= 37.5R_0/C_s$ for QSTK (right) with the ion temperature gradient $a/L_{T_i}=1.21$. The black curves represent the
inner and outer simulation boundaries. We choose the simulation domain upto $\psi_{outer}=0.7\psi_w$ since there is a numerical issue with the EFIT data in QSTK. Also, we choose the plasma profile in a manner so that the mode does not spreads to the boundary of domain.}
\end{figure*}
%
The boundaries of the radial simulation domain are $\psi_{inner}=0.05\psi_w$ and $\psi_{outer}=0.7\psi_w$. The maximum value of the ion temperature gradient length scale measured relative to minor radius is $1.21$ as shown in Fig.~\ref{fig:tiprof}. Other parameters used in the simulation are major radius $R_0=5.58$m, magnetic field on axis $B_0=2.79$T and electron temperature $T_e= 6.50$ keV. After the convergence test, we use $9$ parallel grid points, 121 radial grid points, 4400 poloidal grid points, 200 ions per cell, and $\Delta t= 0.01 R_0/C_s$ where $C_s/R_0=14.11 \times 10^4$sec$^{-1}$ and $C_s=\sqrt{T_e/m_i}$ is the ion acoustic speed. Fig.~\ref{fig:linmodew7x} (left) represents the electrostatic potential of ITG mode on $\zeta=0$ poloidal plane during the linear phase of the nonlinear simulation at $t= 25.0R_0/C_s$. The mode is localized at the outer mid-plane, where the curvature is bad in the toroidal angle with a bean-shape cross section, and it peaks around $\psi$ $\sim$ 0.51$\psi_w$. The mode amplitude peaks at the flux value where the poloidal harmonic number is $m=82$, and the corresponding toroidal harmonic number is $n=71$ with a frequency of $w_r = 1.54 C_s/ R_0 $. The mode propagates in the ion diamagnetic direction having a growth rate of $\gamma = 0.51  C_s/ R_0$, and normalized perpendicular wave number $k_{\perp}\rho_i$=0.55.    

\begin{figure*}
  \centering \includegraphics[width=\textwidth]{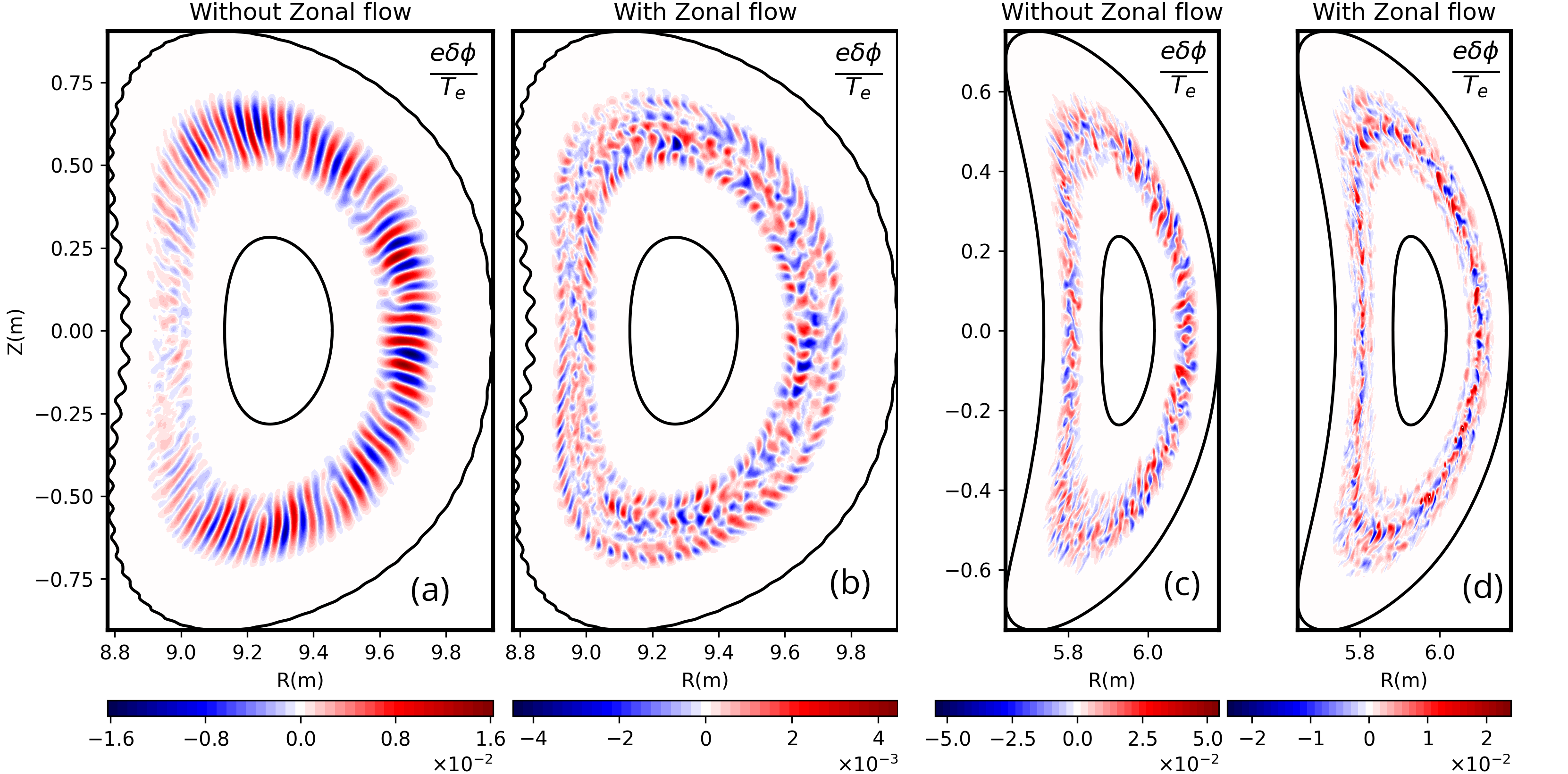}
  \caption{\label{fig:zfmodestel} Contour plots of the electrostatic perturbed potential in the nonlinear phase for both machines with ion temperature gradient $a/L_{T_i}=1.21$. (a) QSTK without ZFs, (b) QSTK with ZFs at $t=55.0 R_0/C_s$, (c) W7-X  without ZFs, (d) W7-X with ZFs at $t=45.0 R_0/C_s$. The black curves indicate the
inner and outer simulation boundaries.} 
\end{figure*}

\begin{figure}
\begin{center}
    
\includegraphics[width=0.5\textwidth]{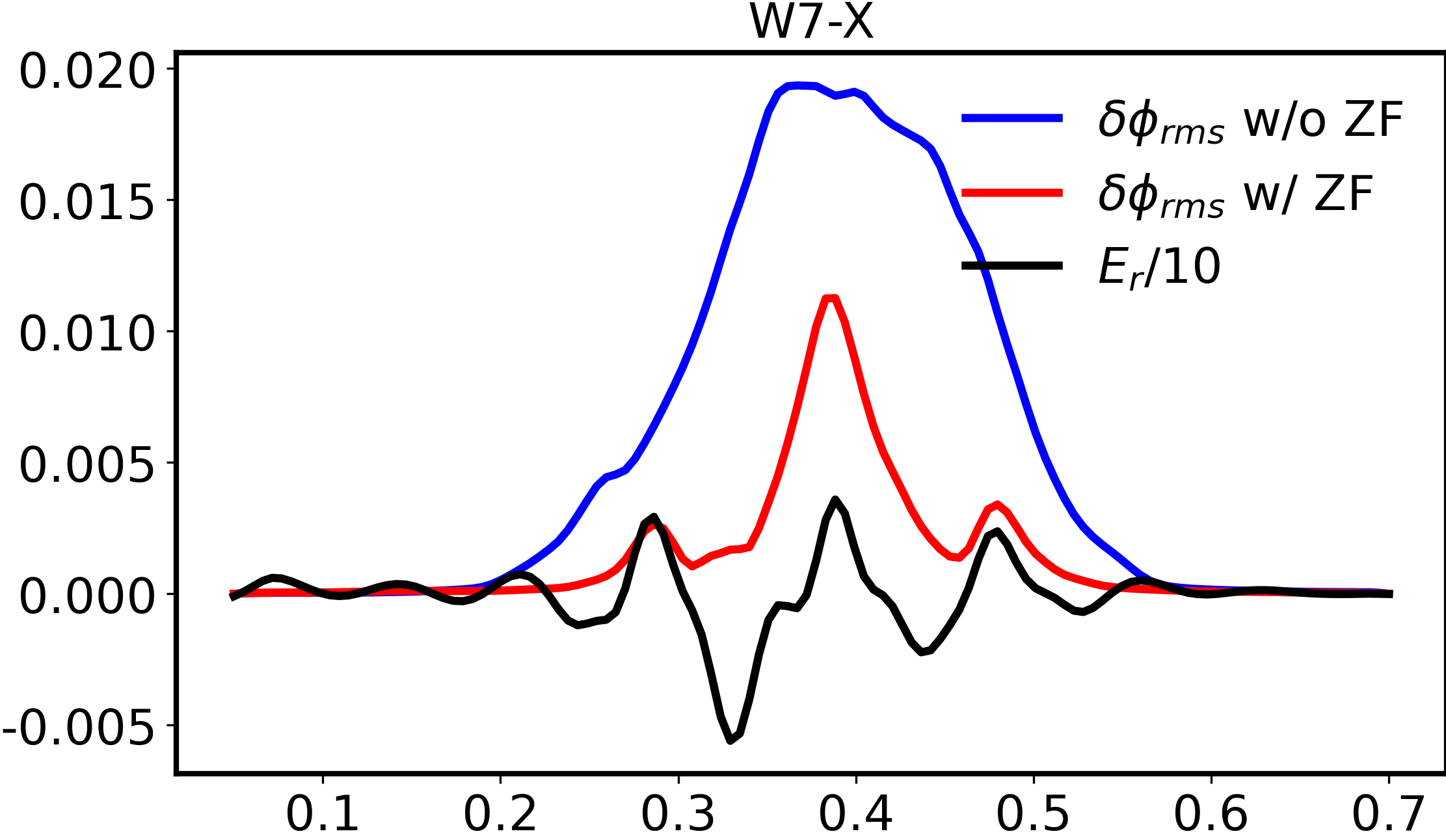}\\
\vspace{1.0cm}
\includegraphics[width=0.5\textwidth]{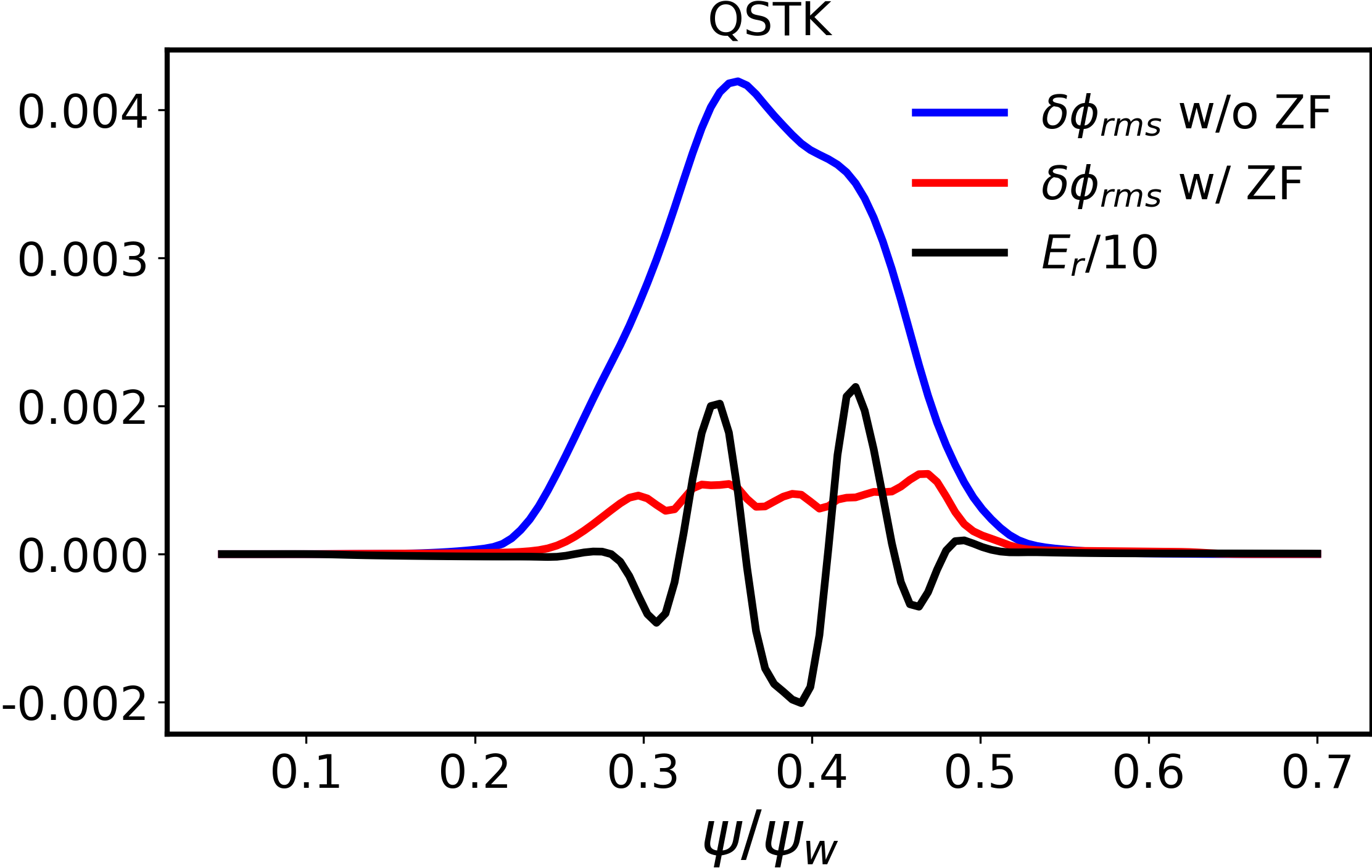}
\caption{\label{fig:phirmsstel} The flux surface variation of root-mean-squared electrostatic perturbed potential ($\delta\phi_{rms}$) with (blue line) and without (red line) zonal flow and the radial electric field ($E_r$) (black line) from the turbulence at the saturation stage of ITG turbulence at time $t=55 R_0/C_s$ for W7-X (top) and $t=65 R_0/C_s$ QSTK (bottom). The electrostatic
potential is normalized with $T_e/e$, and the radial electric field
resulting from the turbulence is normalized with $\sqrt{T_e/e}$. The ion temperature gradient, in this case, is $a/L_{T_i}=1.21$. }

\end{center}
\end{figure}

\begin{figure*}
  \centering \includegraphics[scale=0.45]{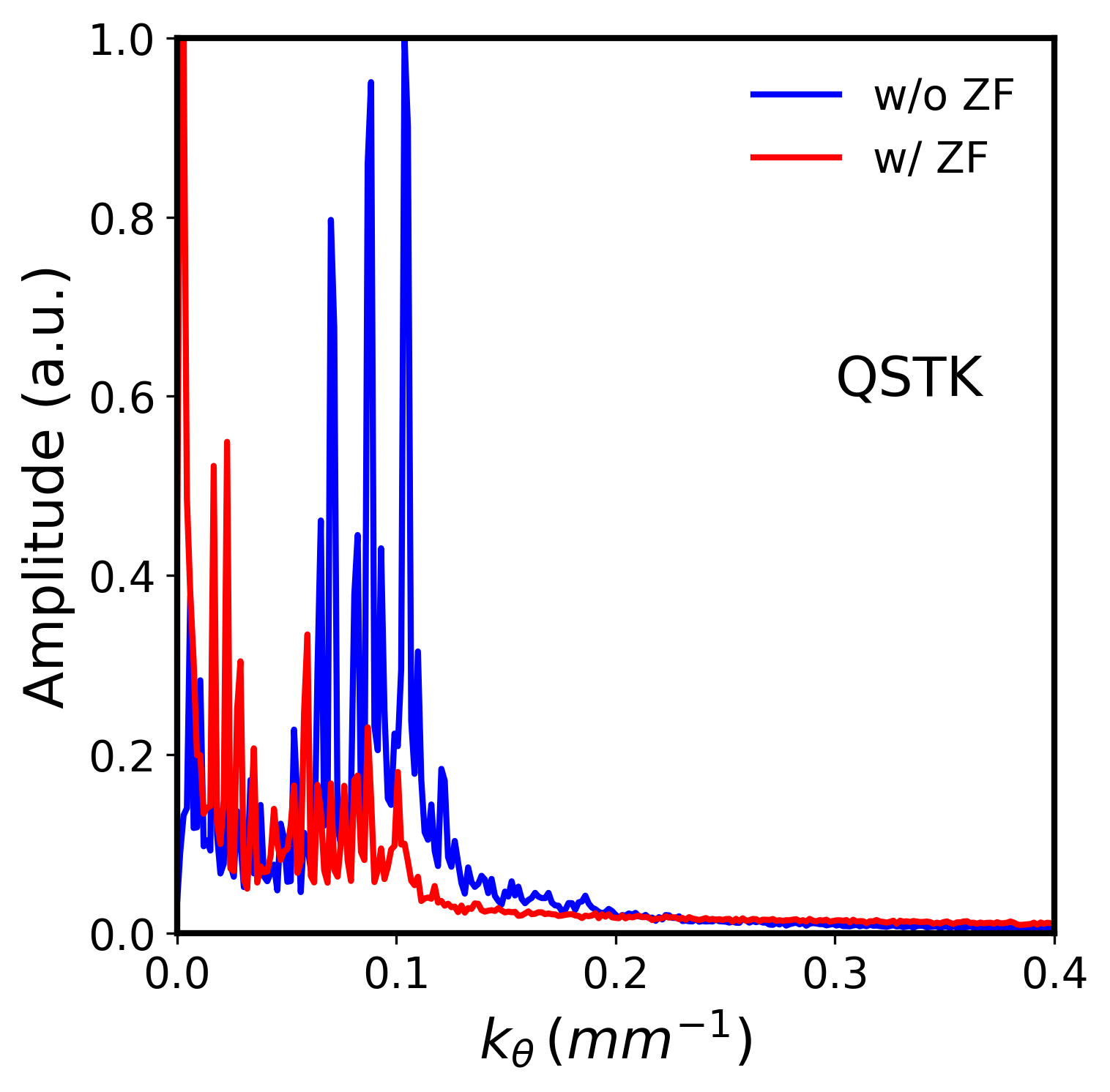}
   \centering \includegraphics[scale=0.45]{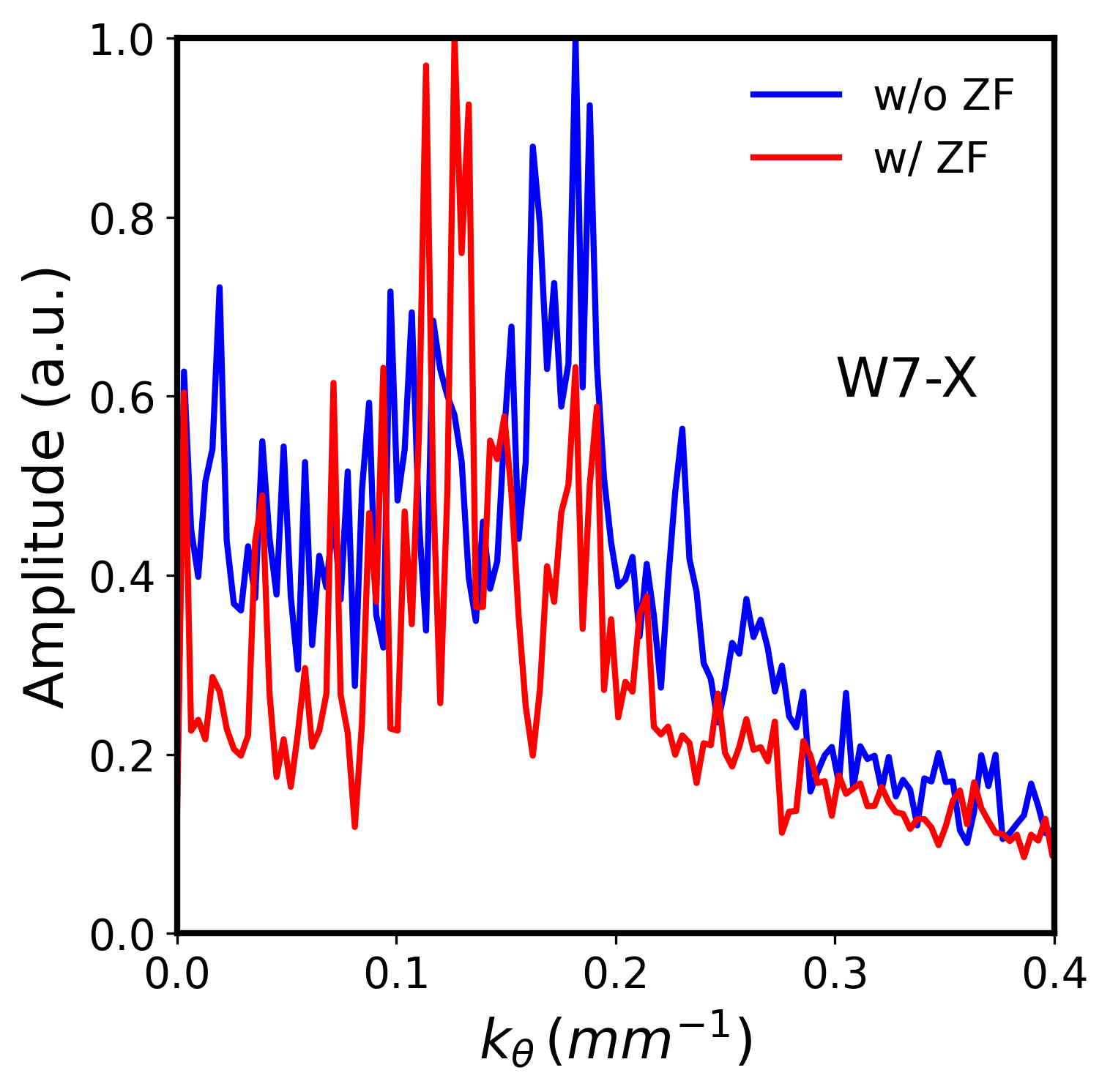}
  \caption{\label{fig:zfpodstel} The poloidal wave number spectrum in QSTK (Left) and W7-X (Right) for the ion temperature gradient $a/L_{T_i}=1.21$. The poloidal wave number decreases for both configurations in the presence of ZFs. For QSTK, the poloidal spectrum is plotted by taking the average over the saturation phase $t=[65.0,72.5]R_0/C_s$ with and without ZFs. For W7-X, the same quantity is plotted by averaging over $t=[32.5,40]R_0/C_s$ with and without ZFs.}
\end{figure*}

\begin{figure*}
  \centering \includegraphics[scale=0.40]{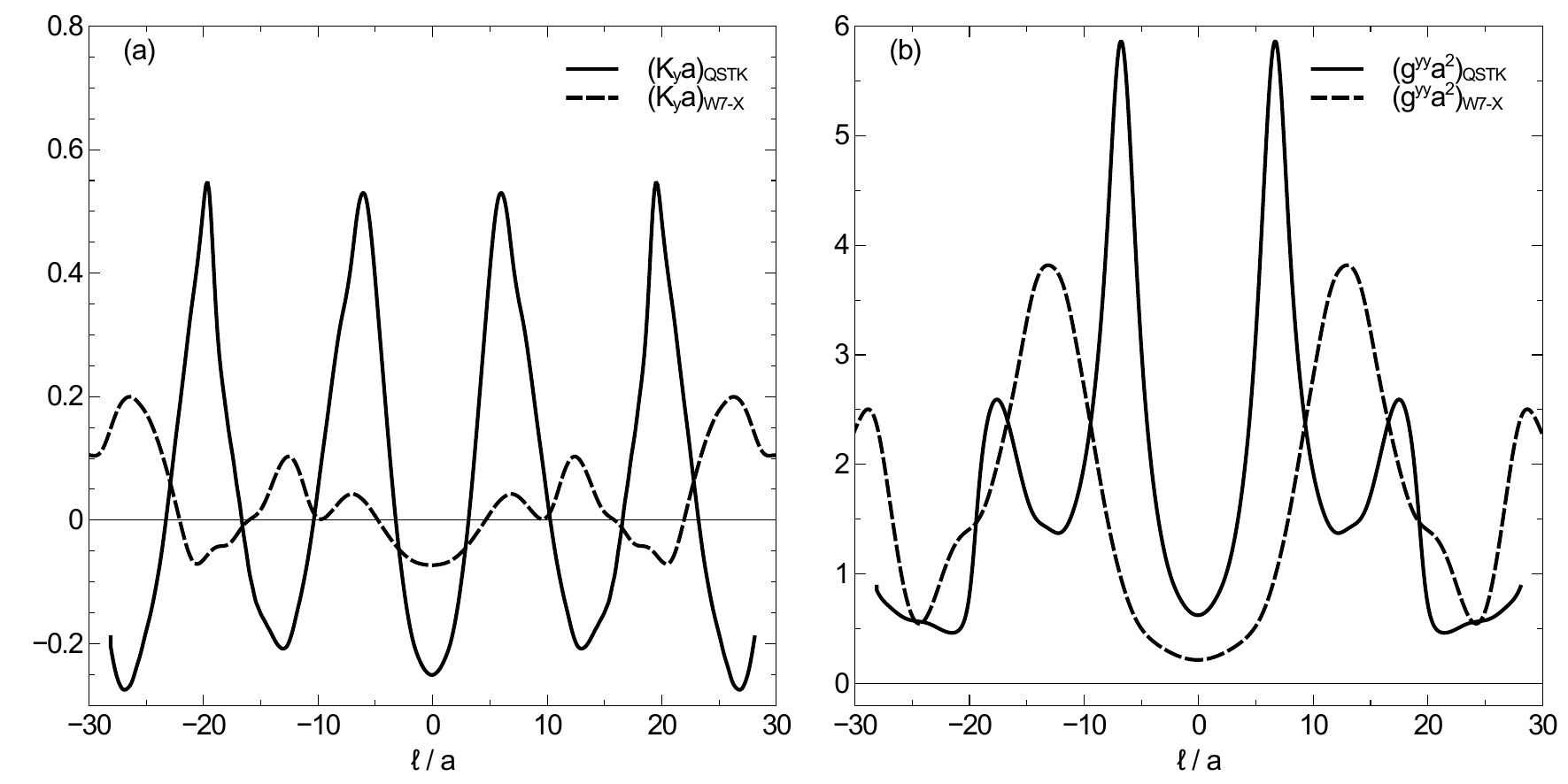}
\begin{tikzpicture}
    \node[anchor=south west,inner sep=0] (image) at (0,0) {%
        \includegraphics[width=0.82\linewidth]{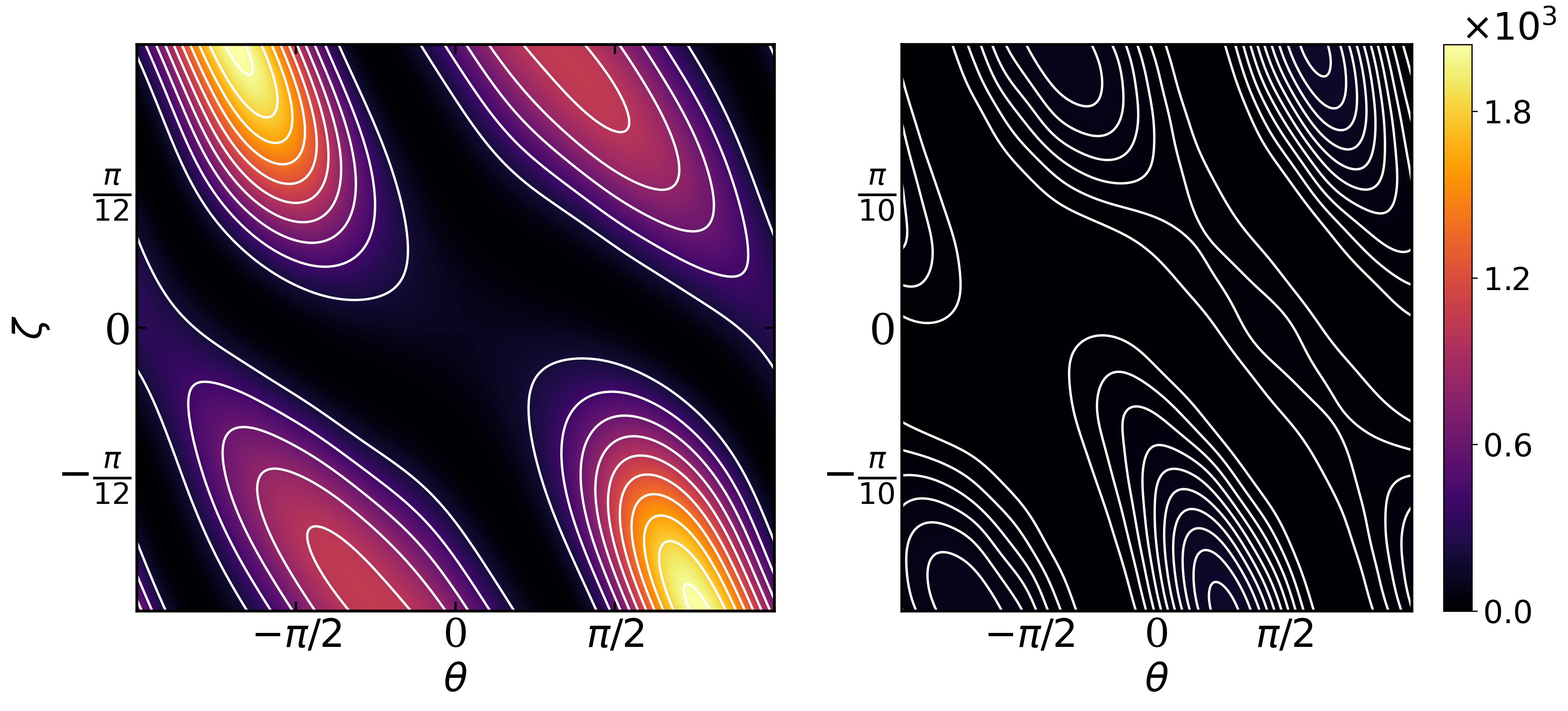}%
    };
    \node[anchor=north west, font=\large\bfseries] at ([xshift=0.02\textwidth, yshift=-0.01\textwidth]image.north west) {(c)};
    \node[anchor=north west, font=\large\bfseries] at ([xshift=0.42\textwidth, yshift=-0.01\textwidth]image.north west) {(d)};
\end{tikzpicture}

\begin{tikzpicture}
    \node[anchor=south west,inner sep=0] (image) at (0,0) {%
        \includegraphics[width=0.82\linewidth]{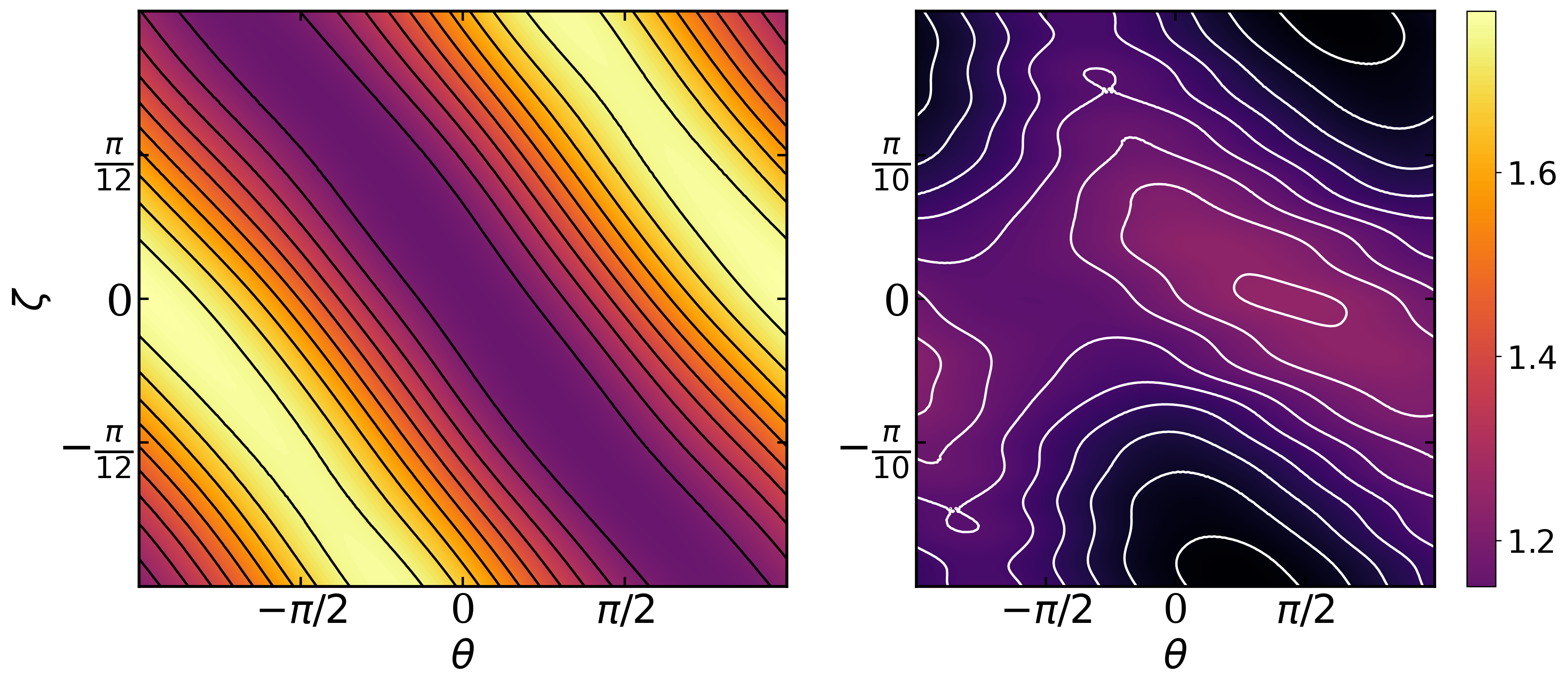}%
    };
    \node[anchor=north west, font=\large\bfseries] at ([xshift=0.02\textwidth, yshift=-0.01\textwidth]image.north west) {(e)};
    \node[anchor=north west, font=\large\bfseries] at ([xshift=0.42\textwidth, yshift=-0.01\textwidth]image.north west) {(f)};
\end{tikzpicture}

  \caption{\label{fig:metrics} Geometric quantities entering the gyrokinetic equation, plotted along the standard flux tube on the outboard midplane at $\zeta=0$, illustrating the difference in drift curvature and squared gradient of the binormal coordinate $\nabla y$ at the radius $\psi / \psi_{\text{edge}}=0.25$, for both QSTK and W7-X. The $y$ coordinate corresponds to the field line label $\alpha$, $\mathbf{B} = \nabla \psi \times \nabla \alpha$, such that $y= \sqrt{\psi_{0}/\psi_{w}}\left(q(\theta - \theta_{0})-\zeta \right)$, with $\psi_{0}$ the chosen flux surface, $q_{0}$ the safety factor at that surface, and $\theta_{0}$ the ballooning angle. The horizontal axes of both plots are in units of arc length normalized to the respective minor radius $a$ of each configuration, while the vertical axes are dimensionless but also normalized to the minor radius of each respective configuration. One toroidal turn is chosen for the extent along the field line in both cases. (a) The drift curvature $K_{y} = (1/B^{2})  \mathbf{B} \times \nabla B \cdot \nabla y$ showing a narrower connection length between ``good'' (positive) and ``bad'' (negative) curvature for QSTK compared to W7-X. (b) Plotting $g^{yy} = |\nabla y|^{2}$ shows that that $|\nabla y|^{2}$ is noticeably larger for QSTK at the central unstable bad curvature well near $\ell=0$, suggesting enhanced finite Larmor radius stabilization at larger poloidal wavenumbers. We also plot the metric quantities using the GTC code. $g_{\theta \theta}$ for (c) QSTK and (d) for W7-X. $g_{\zeta \zeta}$ for (e) QSTK and (f) for W7-X.} 
\end{figure*}

\subsection{\label{subsec: qstkinst} ITG instability in QSTK}

\noindent The linear ITG simulations for QSTK employ identical spatial resolutions and
plasma profiles as those utilized in above analysis for W7-X. The simulation domain is 
restricted to $\psi_{outer}=0.7\psi_w$ due to numerical issues in the QSTK EFIT 
data. The plasma profile parameters are carefully chosen to ensure the mode is localized
within the computational domain, preventing boundary artifacts. The simulation time
step used for the linear simulation is $\Delta t= 0.02 R_0/C_s$ with $C_s/R_0=
9.37\times 10^4$ sec$^{-1}$. Furthermore, the major radius for the QSTK is $R_0=8.40$m,
and the magnetic field on axis value is $B_0=1.01$T. In Fig.~\ref{fig:linmodew7x} 
(right) we show the mode structure of the electrostatic potential of ITG on the 
$\zeta=0$ poloidal plane for QSTK during the growing phase of the nonlinear simulation 
at $t= 37.5R_0/C_s$. The poloidal mode number and the toroidal mode numbers at the 
location where the eigenmode peaks ($\psi$ $\sim$ 0.52$\psi_w$) are $m=59$ and $n=63$,
respectively, with a frequency $w_r =$ 2.64$ C_s/R_0$ propagating in the ion
diamagnetic direction with growth rate $\gamma =$ 0.35 $C_s/ R_0 $ and normalized wave
number $k_{\perp}\rho_i=$0.75. While the normalized temperature gradient $a/L_{T_i}$
  is chosen to be the same in both QSTK and W7-X cases, the absolute ion temperature gradient length scale $L_{T_i}$
  differs due to differences in the local geometry and equilibrium parameters. In both QSTK and W7-X, the linear mode structures of ITG
resemble the typical ballooning structure, localized on the outer midplane, and the 
widths of the linear modes (full width at half maximum) for both stellarators have almost
similar values of $0.06$ in units of $r/a$. At both sides of the radial simulation 
domain, fixed boundary conditions are applied for all fluctuating quantities, and all
the out-of-boundary particles are brought back into the simulation domain through 
energy-conserving boundary conditions and by setting particle weight to be zero.  

\section{\label{sec:nlsimstels} Nonlinear ITG simulations}
\begin{figure}

\hspace{-1.1cm}
\begin{tikzpicture}
    \node[inner sep=0pt] (image) at (0,0) {\includegraphics[scale=0.42]{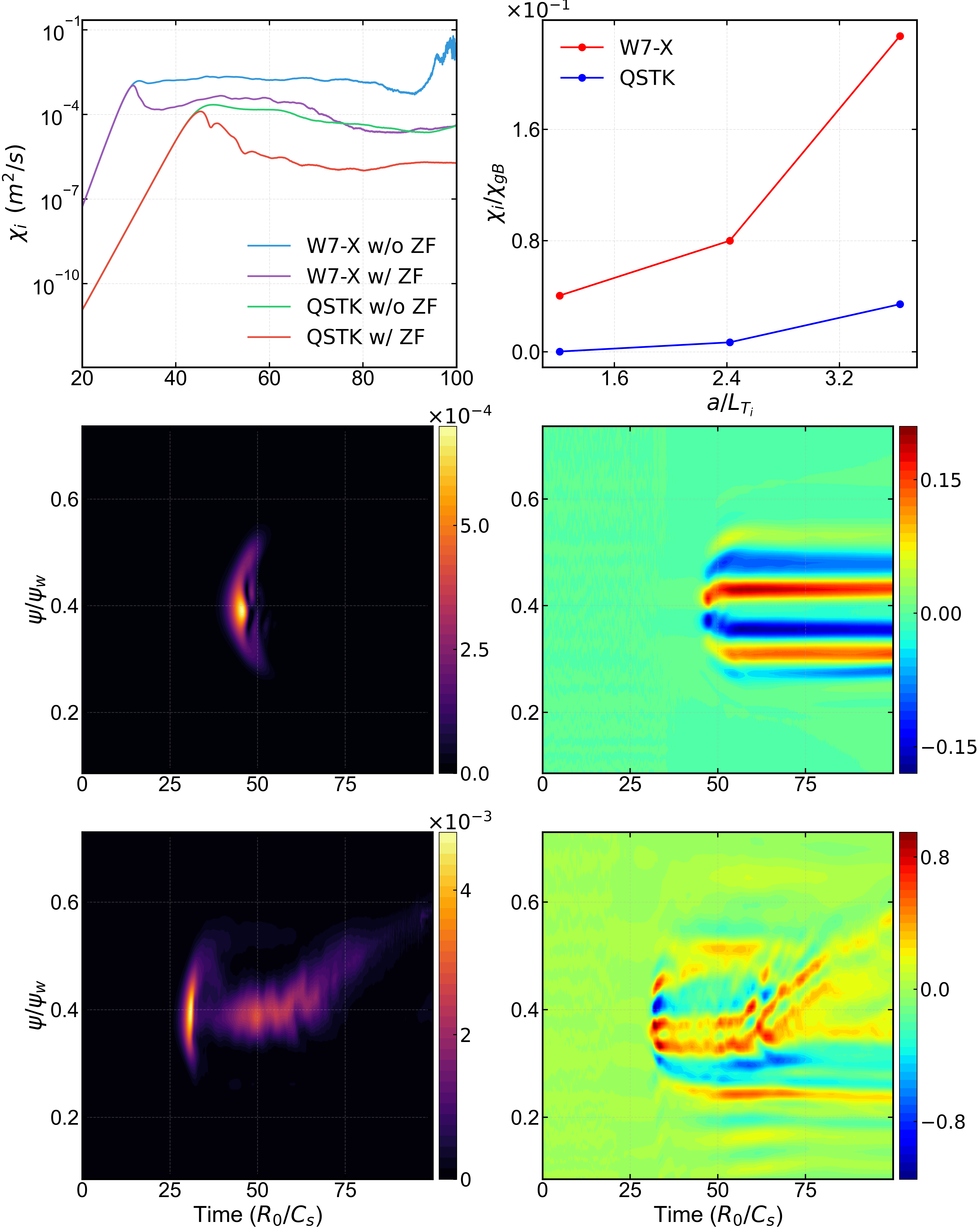}};
    
    \node[anchor=north west, font=\Large\bfseries, black] 
          at ([xshift=2.1cm, yshift=-0.5cm]image.north west) {(a)};
    \node[anchor=north west, font=\Large\bfseries, black] 
          at ([xshift=5.1cm, yshift=-0.5cm]image.north) {(b)};
    
    \node[anchor=north west, font=\Large\bfseries, black] 
          at ([xshift=0.4cm, yshift=3.5cm]image.west |- image.center) {(c)};
    \node[anchor=north west, font=\Large\bfseries, black] 
          at ([xshift=1.2cm, yshift=3.2cm]image.center) {(d)};
    
    \node[anchor=north west, font=\Large\bfseries, black] 
          at ([xshift=0.4cm, yshift=6.8cm]image.south west) {(e)};
    \node[anchor=north west, font=\Large\bfseries, black] 
          at ([xshift=1.2cm, yshift=6.8cm]image.south) {(f)};
\end{tikzpicture}

\caption{\label{fig:chiiw7xqstk}  (a) Comparison of ion heat conductivity ($\chi_i$) in QSTK and W7-X, with and without ZFs for the ion temperature gradient $a/L_{T_i}=1.21$. We find that there is a numerical instability for the case of without ZF in the case of W7-X at late times. (b) In the presence of ZFs, the normalized ion heat diffusivity in gyro-Bohm units is plotted as a function of the ion temperature gradient. We see that the normalized ion heat diffusivity is lower in QSTK compared to the W7-X over the entire range of ion temperature gradient. The time evolution of the radial profile of $\chi_i$ for QSTK and W7-X is shown in (c) and (e), while the corresponding shearing rates are presented in (d) and (f), respectively, for $a/L_{T_i}=1.21$.
}
\end{figure}
\noindent In this section, we focus on the turbulence features for QSTK and W7-X using the same plasma profiles. Specifically, we study the effect of ZFs on the collisionless ITG saturation mechanism in both W7-X and QSTK. The spatial resolutions and the marker particle numbers for these nonlinear simulations are the same as the linear cases; however, the time step for these simulations is $\Delta t=$0.01$R_0/C_s$. Fig.~[\ref{fig:zfmodestel}] represents the contour plots of the electrostatic potential in the nonlinear phase of ITG micro-turbulence in the absence (via numerical suppression) and presence of ZFs for the QSTK and W7-X stellarators. When zonal flows are artificially removed in the nonlinear phase, the linear mode structure spreads radially from the linear eigenmode due to nonlinear toroidal mode coupling Fig.~\ref{fig:zfmodestel}[(a) and (c)]. Once we include ZFs in the simulation, the zonal shear breaks these eddies into fine structures Fig.~\ref{fig:zfmodestel}[(b) and (d)] similar to turbulent self-regulation by ZFs in the tokamak\cite{Lin1998}. To demonstrate the effect of ZFs on the electrostatic potential, we have calculated the root-mean-square value of $\delta\phi$ in the absence and presence of ZFs as the flux surface averaged radial electric field generated by turbulence at the nonlinear stage at $t=55R_0/C_s$, and for QSTK at $t=65R_0/C_s$, see Fig.~(\ref{fig:phirmsstel}). The difference in turbulence potential, shown by the red and blue lines, highlights the suppression of ITG turbulence by ZFs in both configurations. In W7-X, the suppression is $\sim$2.1 times, while in QSTK, it is $\sim$5.9 times, including ZFs, compared to the case without ZFs. This demonstrates the significant role of ZFs in reducing ITG-driven turbulent transport\cite{Singh_2022} in these two optimized stellarators.

\noindent To further evaluate the ZF effect, we have analyzed the poloidal spectrum of the time-averaged electrostatic potential during the nonlinear phase. Fig.~\ref{fig:zfpodstel} shows the time-averaged poloidal wave number spectrum for QSTK (left) and W7-X (right) in the presence and absence of ZFs. For QSTK, we consider the time average from 65.05$R_0/C_s$ to 72.5$R_0/C_s$ and for W7-X, from 47.5 $R_0/C_s$ to 52.5 $R_0/C_s$. The wave number spectra are broad due to the nonlinear mode coupling $k_\theta\in [0,0.15]$ mm$^{-1}$ and $k_\theta\in [0,0.4]$ mm$^{-1}$ for QSTK and W7-X, respectively, in the absence of ZFs. Interestingly, in Fig.~\ref{fig:zfpodstel}, it is shown that the poloidal wave numbers move to a rather low value for QSTK, whereas high poloidal wave numbers still dominate for W7-X. We conjecture that this is a result of the effects of CG optimization, which, by increasing the gradient of the binormal coordinate along magnetic field lines (see Fig. \ref{fig:metrics}), stabilizes ITG modes with large poloidal wavenumbers relative to those with lower wavenumbers. The resulting low wavenumber modes are driven more weakly by toroidal curvature (through the drift factor $\mathbf{v}_{d}$ in the gyrokinetic equation) and are thus expected to have smaller relative growth rates.

\noindent Finally, to quantify the ZF effect on micro-turbulence, we computed the transport coefficients in these two configurations in the presence and absence of ZFs. Fig.~\ref{fig:chiiw7xqstk} (a) shows the time trace of ion heat conductivity, which is calculated in GTC as \cite{Singh_2023}
\begin{eqnarray}
     \chi_i= \frac{1}{\langle |\nabla \psi|^2\rangle n_{i} \frac{\partial T_{i}}{\partial 
    \psi}}  \left\langle \int d^3v \delta f \left(\frac{1}{2} m_{i} v^2 -\frac{3}{2} T_{i} 
    \right) \mathbf{v_E}\cdot \nabla \psi \right\rangle \nonumber
    \label{eq:chiiequation}
\end{eqnarray}

\begin{table}
\caption{\label{tab:comparetable1} Comparison of the effect of ZF for the two stellarators. We compare the ion heat conductivity ($\chi_i$) in the nonlinear regime for the two cases listed below. The reduction ($=\chi_{i_{\text{woZF}}}/\chi_{i_{\text{wzf}}}$) is calculated by taking the ratio of the mean of $\chi_i$ in the nonlinear regime shown in Fig.(~\ref{fig:chiiw7xqstk} a). We also compare the effect of temperature gradient on the reduction of $\chi_i$ in the presence of ZFs in the two machines.}
\begin{tabular*}{\textwidth}{@{}l*{15}{@{\extracolsep{0pt plus
12pt}}l}}
\hline
Case & $a/L_{T_i}$ & Reduction in $\chi_i$ \\
\hline
W7-X (w/o ZF) vs QSTK (w/o ZF) \;\; & $1.21$ & $\sim$ $34.12$  \\
W7-X (w/ ZF) vs QSTK(w/ ZF) \;\;\,& $1.21$ & $\sim$ $27.25$ \\
W7-X (w/ ZF) vs QSTK(w/ ZF) \;\;\,& $2.42$ & $\sim$ $8.6$ \\
W7-X (w/ ZF) vs QSTK(w/ ZF) \;\;\,& $3.63$ & $\sim$ $4.9$ \\
\hline
\end{tabular*}
\end{table}
To calculate the above quantity, we first evaluate the term in numerator and the terms $\langle |\nabla \psi|^2\rangle n_{i}$ in denominator. We then divide the whole quantity by the maximum value of $\partial T_i/\partial \psi$ (as shown in Fig~\ref{fig:tiprof}) to get the value of $\chi_i$. We calculate the reduction due to ZFs by taking the mean value of the $\chi_i$ in the saturated regime as shown in Fig~\ref{fig:chiiw7xqstk}(a). The results for this reduction is presented in Table~\ref{tab:comparetable1}. We also performed a scan in ion temperature gradients ($a/L_{T_i}= [1.21, 2.42, 3.63]$), retaining ZFs, and calculated the normalized ion heat diffusivity in the Fig.~\ref{fig:chiiw7xqstk} (b). QSTK has lower ion heat diffusivity over the entire range of gradients, even above the apparent ITG threshold near $a/L_{T_i} = 1.2$. The more modest (though still significant) suppression factors at higher gradients are expected once the CGs of both configurations are exceeded, as appears to be the case in light of the $a/L_{T}$ scan shown in Fig.~\ref{fig:chiiw7xqstk}(b), since both configurations produce finite heat fluxes at these gradients. In Fig.~\ref{fig:chiiw7xqstk}(c) and Fig.~\ref{fig:chiiw7xqstk}(e), we present the time evolution of the radial profile of $\chi_i$ for QSTK and W7-X, respectively. Similarly, in Fig.~\ref{fig:chiiw7xqstk}(d) and Fig.~\ref{fig:chiiw7xqstk}(f), we plot the corresponding shearing rate defined as $\omega_E= (\partial^2\langle\phi\rangle/\partial \psi^2) (\Delta r/\Delta \theta) R B_{\theta}/q$\cite{hahm1995flow} where, $\Delta r$  and $r\Delta \theta$ are the radial and poloidal correlation lengths, respectively, and $B_{\theta}$ is the poloidal magnetic field. We assume that the radial and poloidal correlation lengths are equal for the purpose of evaluating $\omega_E$. We observe that W7-X exhibits a higher shearing rate than QSTK. The zonal flow generated during nonlinear ITG saturation is rapidly damped by collisionless magnetic pumping effects ~\cite{PhysRevLett.80.724}, resulting in a lower residual level. Linear GTC simulations indicate higher residual levels for QSTK (0.48) compared to W7-X (0.27). A key feature of Fig. 9 (c-f) is that the nonlinear frequency of zonal flow in W7-X is higher than in QSTK. Additionally, the radial structure of QSTK is more coherent and stable than that of W7-X, suggesting a stronger nonlinear instability of zonal flows in W7-X.

\section{\label{sec: conclude} Conclusion and Discussion}

In this work we have carried out the study of ITG-driven turbulence in the optimized stellarators W7-X and QSTK.  The latter design resulted from a recent optimization study that targeted the critical gradient of the ITG mode \cite{PhysRevResearch.5.L032030}.  We found a sensitivity of the turbulence saturation level on the zonal flows. The  ion heat flux, differed by a large factor ($\sim 34$) between the two stellarators at the lowest gradient, where QSTK is close to the ITG marginality. Such a large relative factor at this gradient suggests a threshold behavior in line with the targeting of a high linear critical gradient for ITG modes in QSTK. At higher temperature gradients, apparently above this threshold, QSTK continues to enjoy lower nonlinear heat fluxes in comparison to W7-X, perhaps in part because of reduced linear growth rates for ITG modes. We thus expect CG optimization to continue to be useful in guiding stellarator design for reduced ion transport, whether as a result of improved thresholds at low gradients, lower growth rates at high gradients, or through some interplay of the two effects.  ZF generation is more pronounced when ITG turbulence is near marginal stability, i.e., when the linear growth rate is low. In such conditions, zonal flows can persist longer and effectively suppress turbulent transport, as illustrated in Fig. 9. This study provides valuable insights into how 3D geometries, such as QSTK and W7-X stellarators offer a crucial tool for designing and optimising new stellarators. However, the ultimate determinant of a stellarator’s feasibility as a fusion reactor will be the level of turbulent transport, with the ability to self-regulate playing a pivotal role. The present focus of this paper is to study the zonal flow physics with adiabatic electrons. However, the kinetic electrons will play a significant role in evaluating more accurate heat flux as described in detail in \cite{Singh_2022,Garcia, Zocco}, which will be carried out in future work.

\section*{Acknowledgments}
This work is supported by Board of Research in Nuclear Sciences (BRNS Sanctioned no. and 57/14/04/2022-BRNS), Science and Engineering Research Board EMEQ program (SERB sanctioned no. EEQ/2022/000144), National Supercomputing Mission (NSM), US Department of Energy under Award No.DE-SC0024548 and DE-FG02-07ER54916. We acknowledge National Supercomputing Mission (NSM) for providing computing resources of `PARAM PRAVEGA’ at S.E.R.C. Building, IISc Main Campus Bangalore, which is implemented by C-DAC and supported by the Ministry of Electronics and Information Technology (MeitY) and Department of Science and Technology (DST), Government of India,  ANTYA cluster at Institute of Plasma Research, Gujarat, and by US DOE SciDAC and INCITE. A.T. thanks the University Grants Commission (UGC) for supporting him as a Senior Research Fellow (SRF).



\bibliography{3_Untracked_mybib}

\vspace{0.5cm}
\begin{center}
\line(1,0){200}
\end{center}

\end{document}